\documentclass[prb,
twocolumn,
showpacs,floatfix
]{revtex4}
\usepackage{graphicx,amsmath,amssymb,latexsym,amsfonts,subfig
ure,bbm}
\usepackage[usenames]{color}
\usepackage{epic}
\usepackage{epsfig}
\newcommand{\x}{\mathbf{x}}
\newcommand{\EF}{E_{\rm F}}
\newcommand{\kf}{k_{\rm F}}

\newcommand{\Arc}{\arccos(\varepsilon/\Delta)}

\newcommand{\delN}{\delta\! \rho}
\newcommand{\dely}{\delta\! y}
\newcommand{\lamF}{\lambda_{\rm F}}
\newcommand{\delyAv}{\overline{\delta\!y}}
\newcommand{\vecUV}{\left(\begin{array}[c]{c}u(\x)\\v(\x)\end
{array}\right)}

\begin{document}
\title{Non-retracing orbits in Andreev billiards}
\author{Florian Libisch}
\email{florian@concord.itp.tuwien.ac.at}
\author{Stefan Rotter}
\author{Joachim Burgd\"orfer}
\affiliation{Institute for Theoretical Physics, Vienna University of 
Technology\\Wiedner Hauptstra\ss e 8-10/136, A-1040 Vienna, Austria, European
Union}
\date{\today}

\begin{abstract}
The validity of the retracing approximation in the semiclassical quantization
of Andreev billiards is investigated. 
The exact energy spectrum and the eigenstates of normal-conducting,
ballistic quantum dots in contact
with a superconductor are calculated by solving the Bogoliubov-de Gennes 
equation and compared with the semiclassical Bohr-Sommerfeld quantization for
periodic orbits which result from Andreev reflections.
We find deviations that are due to the assumption of exact 
retracing electron-hole orbits rather than the semiclassical approximation, as a 
concurrently performed Einstein-Brillouin-Keller
quantization demonstrates. We identify three different mechanisms
producing non-retracing orbits which are directly identified through
differences between electron and hole wave functions.
\end{abstract}



\maketitle

\section{Introduction}
\begin{figure}[tb]

\epsfig{file = 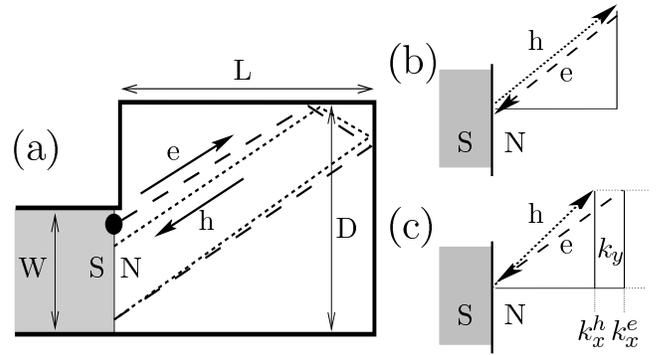,width=8.5cm}
\caption{(a) Andreev billiard with a rectangular normal-conducting 
($N$) region. 
The superconducting ($S$) lead is assumed to be half-infinite (shaded
area). The dashed and dotted lines 
depict an ``almost retracing'' electron-hole orbit created by Andreev
reflection at the SN-interface. For better visibility the starting point of
the orbit is marked by a dot. (b) perfect Andreev reflection, 
i.e.~$\theta_e = \theta_h$. (c) imperfect Andreev reflection,
$\theta_e\ne\theta_h$, $k_y^e = k_y^h$.}
\label{fig:dot}
\end{figure}

The dynamics of hybrid systems consisting of a normal-conducting (N) quantum
dot and a  superconducting (S) waveguide have recently raised much
experimental~\cite{Morpurgo,Pothier,Charlat} and
theoretical~\cite{Colin-review, Bagwell, StoneAnd, BenRandMat, MagAndreev,
PseudMagAndreev,Gyorfy,Mortensen,JozNegLen} interest.
This is due to the unusual and sometimes counterintuitive 
properties of the interface of normal-- and superconducting regions.
An electron moving inside the N region will be
coherently scattered into a hole upon contact 
with the superconductor [see Fig.~\ref{fig:dot}(a)].\cite{BTK} 
This phenomenon is commonly referred to as Andreev 
reflection.\cite{Andreev} In the frequently used and remarkably 
successful semiclassical Bohr-Sommerfeld (BS)
method\cite{Ihra,JozBoxDisk,JozCakeChaos} it is assumed that the 
Andreev reflection is perfect, i.e.~that the path of
the backscattered hole will exactly trace that of the incoming electron
($\vec{v}_e=-\vec{v}_h$) [see Fig.~\ref{fig:dot}(b)]. 
The consequences of this assumption
are profound: all trajectories emanating from the SN-interface are strictly
periodic. The classical dynamics of the combined SN-system become entirely
regular, even and in particular when the normal conducting cavity would
feature hard chaos.\cite{oppen} 
Unlike chaotic or integrable systems, periodic orbits are
no longer isolated but form continuous manifolds that dominate the classical
phase space and, in turn, the density of states (DOS).\cite{DeGennes,Schomerus,Zirnbauer}
The BS approach\cite{Ihra, JozBoxDisk,
JozLogCont} to the DOS of an Andreev billiard relies on three 
assumptions: exact retracing of electron-hole
trajectories (referred to as assumption A1 in the following), 
the absence of any 
quasi-periodic orbits other than the ones caused 
by Andreev reflection (assumption A2), and 
the applicability of semiclassical approximations (assumption A3),
i.e.~$\lambda_D\ll\sqrt{A}$, where $\lambda_D$ is the de Broglie wavelength
and $A$ is the area of the $N$ billiard.

The BS approximation has been found
to be suprisingly accurate,\cite{JozBoxDisk,JozLogCont}
 even in the presence of a magnetic field
\cite{Ihra} or a soft wall potential.\cite{Lib1} 
However, some difficulties have emerged:
for example, the excitation gap in the DOS for billiards with a chaotic $N$
cavity cannot be properly accounted for.\cite{Vavilov,Altland,Melsen} 
A direct inspection of
the electron- and hole-components of the wave functions\cite{libThesis,Lib1} revealed
markedly different patterns, clearly signaling a breakdown of the retracing
approximation.

In the present communication, we investigate the role of non-retracing orbits
in Andreev billiards. Non-retracing electron-hole orbits leave their
mark on the
Andreev wave functions which we determine by solving the Bogoliubov-de
Gennes equation with the help of the modular recursive
Green's function method (MRGM).\cite{StephanDipl,Stephan}
We find that Andreev states which correspond to non-retracing orbits 
break the close correspondence between the electron and hole wave
function patterns.
The analysis of the eigenstates allows us to check the merits and limitations
of the BS approximation against the exact quantum mechanical
approach. We identify different sources for the breakdown and explore separately each
of the assumptions A1 to A3 listed above.

This paper is organized as follows. In Sec.~\ref{SecMeth} the method for
solving the quantum mechanical eigenvalue problem
as well as the Bohr-Sommerfeld approach to Andreev billiards is briefly 
reviewed. In section \ref{validity}
we identify different mechanisms leading to non-retracing orbits in
Andreev billiards. The paper concludes with a short summary in
Sec. \ref{SecSum}.

\section{Model system}\label{SecMeth}
We consider as a model system a rectangular normal conducting cavity
of width $D$, length $L$, and area $A=LD$ 
which is attached to a half-infinite lead of width
$W$ [see Fig.~\ref{fig:dot}(a)]. 
For technical reasons we focus on an $N$-cavity
that features classically regular motion. 
Since within the framework of perfect Andreev
reflections combined SN-Andreev systems become regular and, in fact, periodic
irrespective of the underlying regular or chaotic dynamics of the $N$-cavity,
we expect many of our results to be valid for arbitrary
$N$-cavities. As we consider only one superconducting lead, we choose
the gap parameter $\Delta$ to be real.
The superconducting coherence length ($\xi$) is assumed to be
small compared to the linear dimension of the rectangular cavity
($\sim\!\!\sqrt{A}$). Under that assumption we may use a step function model
\cite{BenRandMat} at the SN-interface, $\Delta = |\Delta_0|\
\theta(x_{NS}-x)$. 
This SN-interface, which is located at $x_{NS}=0$, is
assumed to be ideal, meaning that there is no tunnel barrier
and no mismatch between Fermi energies and effective masses 
in the N- and S-region. An exact treatment of the Bogoliubov-de Gennes 
equation reveals that this results in a 
probability for Andreev reflection close to
unity (up to corrections of the order $\Delta^2/\EF^2$) in the energy range we consider, 
$0 \le \varepsilon \le \Delta = 0.02 \EF$.\cite{Mortensen,BTK} 
Furthermore, the
phase coherence length of the scattering process is assumed to be infinitely
long, resulting in fully ballistic dynamics.

\subsection{Quantum mechanical solution}
We describe the S-N hybrid system by the Bogoliubov-de Gennes (BdG)
equation \cite{DeGennesArt}
\begin{equation}
\label{Meth:BDG}
\left(\begin{array}[c]{cc}H_0&\Delta\\
\Delta^*&-H_0^*\end{array}\right)\vecUV=\varepsilon\vecUV.
\end{equation}
 $H_0 = p^2/(2m_{\rm{eff}})+V(r)-E_F$ is the single particle Hamiltonian. The
 electron (hole) quasiparticle wavefunctions are denoted by $u$ ($v$),
 $\varepsilon$ is
the excitation energy of the electron (hole) above (below) the Fermi energy
$E_F$. Throughout this publication we consider only bound Andreev states for
which the solutions of (\ref{Meth:BDG})
in the superconducting region are exponentially decaying. To identify
the bound Andreev states, we use a wave function
matching technique\cite{JozBoxDisk} in combination with the Modular Recursive
Green's function method\cite{StephanDipl,Stephan} as outlined in
Ref.~\onlinecite{Lib1}.\\Note that the above procedure is exact in the sense
that it does not rely on the usual Andreev appoximation,
namely $\Delta \ll E_F$, and the assumption of
approximately perpendicular angles of incidence on the 
SN-interface.\cite{BTK,Andreev}
This feature will be important when we consider the effect of imperfect 
Andreev reflections on the quantum and semiclassical spectra.

\subsection{Semiclassical analysis}
The intuitive picture of ideal Andreev reflections 
[Fig.~\ref{fig:dot}(b)] lends itself to a
semiclassical description and, more specifically, to a periodic-orbit
quantization. An electron approaches the SN-interface with wave number 
\begin{equation}\label{eq:2a}
k_e=\sqrt{2(E_F+\varepsilon)}
\end{equation}
and angle $\theta_e$ relative to the interface normal. The hole leaves the
interface with wavenumber 
\begin{equation}\label{eq:2b}
k_h=\sqrt{2(E_F-\varepsilon)}\,.
\end{equation}
and a corresponding angle $\theta_h$.
Since the component along the interface is exactly conserved 
because of translational
invariance, $k_{e,y}=k_{h,y}$ (or $v_{e,y}=-v_{h,y}$) the 
components normal to the interface will be, in general, 
different $k_{e,x}\neq k_{h,x}$, leading to
imperfect retracing [Fig.~\ref{fig:dot}(c)]. 
Only in the limit $\varepsilon\rightarrow 0$
(or $\varepsilon/E_F\rightarrow 0$) perfect Andreev reflection 
$k_{e,x}=k_{h,x}$ (or $v_{e,x}=-v_{h,x}$) ensues. The retracing approximation
consists now of the assumption A1 
of perfect reflection, $v_{e,x}=-v_{h,x}$, for all
$\varepsilon$ in the interval $0\leq\varepsilon\leq\Delta$.
The validity and breakdown of this assumption will be explored in the
following. 

Under assumption A1 all trajectories emanating from the
SN-interface are strictly periodic. For $N$-cavities featuring hard chaos,
every trajectory will eventually hit the SN-interface, thus yielding a globally
periodic system. For $N$-cavities with mixed or regular dynamics, certain regions
of the phase space may remain decoupled from the interface and thus may
feature both continuous manifolds of strictly periodic orbits and islands with
quasi-periodic motion. Neglecting the latter contribution to the DOS
implies assumption A2.

The BS quantization of the continuum of perfectly periodic orbits visiting the
SN-interface yields for the action $S$ of periodic orbits created by pairs of
electron-hole trajectories (with wavenumbers $k_e$, $k_h$),\cite{Ihra}
\begin{equation}\label{BScon}
S = l(k_e-k_h) = 2\pi n + 2\arccos(\varepsilon/\Delta).
\end{equation}
The variable $l$ stands for the length
of an arbitrary path connecting one point at the SN-interface with another.  
The quantum number $n=0,1,2,\ldots$ accounts for the quantized
difference in action between the electron and the hole 
part of the periodic orbit. The second term on the right hand side of
Eq.~(\ref{BScon}), $\Arc$, is the Maslov index\cite{Brack} for one reflection
at the SN-interface. Including this energy-dependent phase shift at the point
of Andreev reflection extends\cite{JozBoxDisk} 
the accuracy of the BS description to energy 
ranges $0\le\varepsilon\le\Delta$\,. 
An even more accurate semiclassical description of the Andreev reflection 
process will be discussed in a subsequent publication.\cite{Lib3}
%
 
The semiclassical BS approximation for the 
state counting function can then be written as,\cite{Ihra}
\begin{equation}\label{NBS}
  N_{BS}=M\sum_{n=0}^\infty\int_{l_n(\varepsilon)}^\infty P(l) dl\,,
\end{equation}
where $l_n(\varepsilon)$ is defined by
\begin{equation}
l_n(\varepsilon) = 
\left[n\pi + \rm{arccos}\left(\frac{\varepsilon}{\Delta}\right)
\right]\frac{ k_{\rm{F}}}{\varepsilon}\,.\label{epsqn} 
\end{equation}
The geometry of the cavity enters via the classical path length distribution 
$P(l)$. The quantity $P(l)\,dl$ is the classical probability that an electron
entering the $N$ region at the SN-interface returns to the interface after a
path length in the interval $[l,l+dl]$.

In the special case that the Andreev system becomes separable, in particular
for the geometry $W=D$ [see the inset in Fig.~\ref{fig2}(a), and Fig.~\ref{fig2geom}(a)], 
an alternative semiclassical
approximation, the Einstein-Brillouin-Keller (EBK) quantization, becomes 
applicable. This alternative is of particular interest as it only relies on
assumption A3, but does not involve the additional assumptions A1 and
A2, thus allowing to disentangle the validity of
the semiclassical approximation from that of the retracing approximation.
For an EBK quantization of the cavity with $W=D$ 
the separability yields two quantization conditions, 
the first one of which pertains to the 
motion in $x$ direction [compare with
Eq.~(\ref{BScon})],
\begin{equation}
   \oint_{\mathcal C_x} k_x^{e,h}{\rm d}x = 2L(k_x^e - k_x^h)
= 2\left[n\pi + \arccos\left(\frac{\varepsilon}{\Delta}\right)\right]\,.\label{cx}
\end{equation}
The second quantization condition in $y$ direction reads
\begin{equation}
   \oint_{\mathcal C_y} k_y{\rm d}y = 2 k_y D = 2 m \pi\,.\label{cy}
\end{equation}
The contour $\mathcal C_x$ ($\mathcal C_{y}$) stands for a path
along a horizontal (vertical) line in the normal conducting billiard.
Introducing the short hand notation $k_x^m(\pm\varepsilon)=k_m^\pm$ 
with
\begin{equation}
k_m^\pm = \sqrt{2(\EF\pm\varepsilon) + \left(\frac{m\pi}{W}\right)^2}\,,
\end{equation}
the quantization condition for $x$ [Eq.~(\ref{cx})] reads
\begin{eqnarray}
  \left[n\pi + \arccos\left(\frac{\varepsilon}{\Delta}\right)\right] 
  = L\left[k_m^+-k_m^-\right]\,.
  \label{BSQuan}
\end{eqnarray} 
This is a transcendental equation in $\varepsilon(n,m)$, which allows 
us to calculate individual energy levels in a
cavity with $W=D$ semiclassically. Note that in contrast to this EBK result,
the
BS approach yields only an approximation to the smoothed state counting
function $N(\varepsilon)$. We therefore transform the EBK results for
$\varepsilon(n,m)$ into a state counting function 
\begin{equation}\label{eq:ebktransform}
N_{EBK}=\int_0^{\varepsilon}d\varepsilon'\sum_{m,n}
\delta\left[\varepsilon'-\varepsilon(n,m)\right]
\end{equation}
and compare both
semiclassical approximations with the quantum results. 
EBK quantization of Andreev billiards can be viewed as the analogue to
the adiabatic quantization of smooth soft-wall chaotic billiards, for
completely integrable systems.\cite{Silvestrov,Goorden} 
As the EBK quantization invokes quasi-periodic
rather than periodic motion no assumption of retracing is involved.

\section{Validity of the retracing approximation}\label{validity}
We will now explore the validity and breakdown of the retracing approximation
and the resulting BS quantization by considering rectangular cavities 
(Fig.~\ref{fig2geom})
with different ratios $D/W$ and $D/L$. With this parameter space at our
disposal we can probe and disentangle the validity of assumptions A1 to A3.

\subsection{Quadratic $N$ cavity}

\begin{figure}[hbt]
\epsfig{file=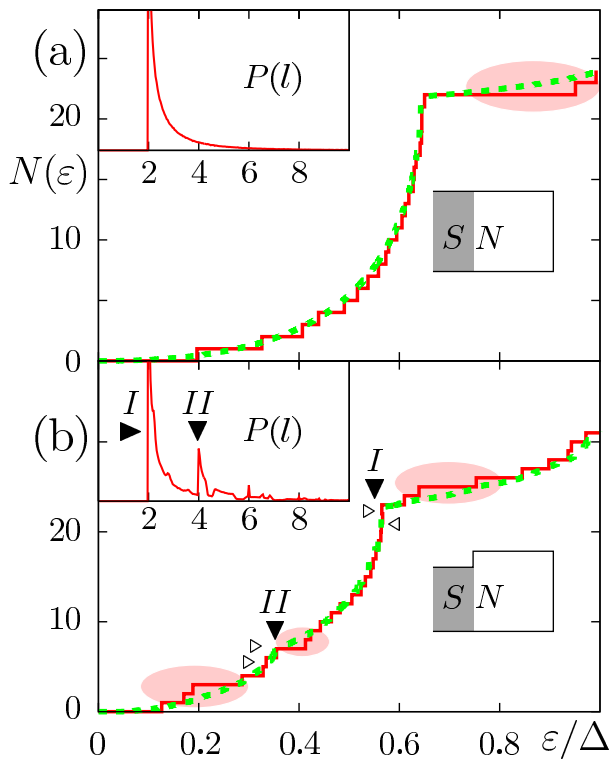,width=8cm}
\caption{(Color online) The quantum mechanical state counting 
function $N_{\rm QM}(\varepsilon)$ (solid red staircase) 
and its semiclassical BS estimate $N_{\rm BS}(\varepsilon)$ (dashed green line)
for two quadratic cavities with different lead widths 
as shown in the insets [$\kf = 20.5\pi/W$, $W=L($a$),\;0.8L($b$)]$. The top left insets show
the classical path length distribution in units of the cavity length $L$. 
Solid triangles mark pronounced cusps in $N(\varepsilon)$ and their 
classical origin. The quantum number 
$n$ (Eq.~\ref{BScon}) increases from 0 to 1 at $I$. Open triangles mark
the energy positions of states whose wavefunctions are displayed 
in Fig.~\ref{fig1}. Shading marks the regions where the BS 
approximation begins to deviate from quantum results.}
\label{fig2}
\end{figure}
\begin{figure}[hbt]
\epsfig{file=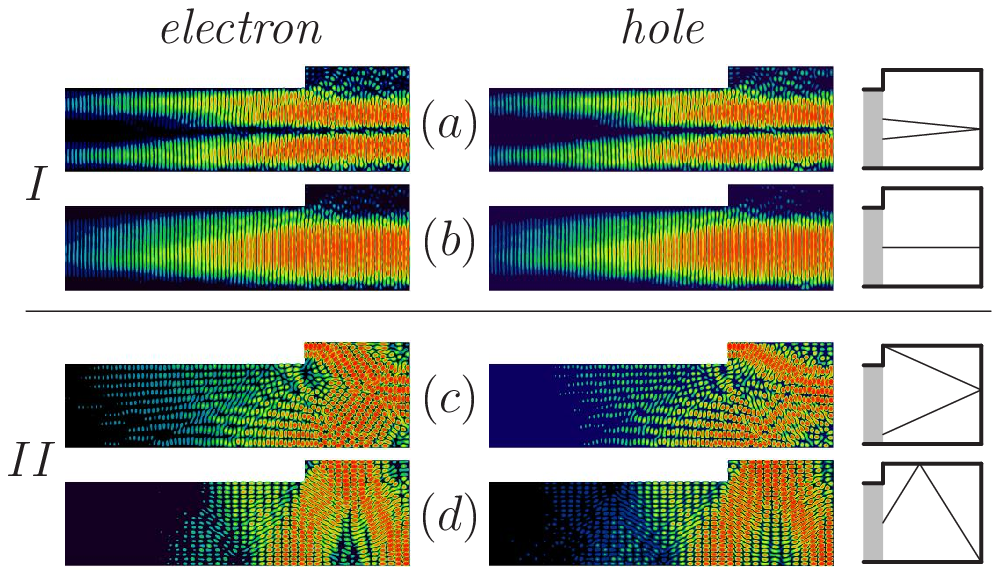,width=8.5cm}
\caption{(Color online) $(a)-(d)$ shows the electron and hole wavefunctions
  $|u(\x)|^2$ and $|v(\x)|^2$ at values of $\varepsilon$ indicated by open
  triangles in Fig. \ref{fig2}(b): $\varepsilon/\Delta$ is 
  $(a)$ 0.565, $(b)$ 0.566, $(c)$ 0.33, $(d)$ 0.355. 
The corresponding classical orbits are indicated on the right.}
\label{fig1}
\end{figure}

We consider first the quadratic cavity with $D=L$. Here two cases arise: for
$D/W=1$, the Andreev billiard becomes separable while for $D/W>1$ the Andreev
system is non-separable even though the $N$ cavity itself 
is separable. In the latter case the
SN-boundary condition breaks the separability. The location of the
placement of the SN-interface
for $D/W>1$ provides an additional degree of freedom for our investigation to
be exploited below. A comparison between the BS approximation and the exact
quantum mechanical calculation for the counting function $N(\varepsilon)$ is
shown in Fig.~\ref{fig2}. Overall, for $D/W=1$ (a) and $D/W=1.25$ (b), the BS
approximation performs quite well. In particular, it reproduces and 
intuitively explains the cusps seen in $N(\varepsilon)$. 
They are due to the sharp
peaks in the pathlength spectrum. The position of the cusps is predicted very
accurately by the BS approximation 
with an error well below the mean level spacing of $0.09\Delta$.

Assuming the
validity of the retracing approximation, the
semiclassical quantization condition [Eq.~(\ref{epsqn})] allows to map every
excitation energy $\varepsilon$ onto a path length $l_n$ of a periodic Andreev
orbit. Consider, e.g., the energy at
the cusp marked $I$ in Fig.~\ref{fig2}(b) that corresponds to a path length
$l_0 =2L$, which is the length of the
shortest classical Andreev-reflected orbit of the system [see $P(l)$ in the
inset of Fig.~\ref{fig2}(b)]: The electron leaves
perpendicular to the SN-interface, 
is reflected at the opposing wall and returns to the SN-interface. 
Quantum mechanical wave functions evaluated at the cusp energy
indeed feature a pronounced enhancement along the
orbit ``bundle''\cite{LudgerBundle} with length $l=2L$ 
[see in Fig.~\ref{fig1}(b)]. Wavefunctions with neigboring energy values below
the cusp feature additional nodes in transverse direction 
[see Fig.~\ref{fig1}(a)]. Note that also eigenstates near the cusp marked
$II$ in Fig.~\ref{fig1}(c, d) correspond very nicely to the path-bundles of the
expected length. Consistent with the good agreement of the BS 
quantization with the quantum calculations,
the electron and hole wavefunction densities agree very well with each other.
To the extent that bundles of classical trajectories cause the
density enhancement in the wavefunction, and bundles of hole- and
particle-orbits agree with each other, this is to be expected. Conversely, a
hallmark of the breakdown of retracing are distinct density ditributions in
the particle ($u$) and hole ($v$) wave functions, as has been recently
observed.\cite{libThesis,Lib1} Looking more closely, we find 
discrepancies between the exact
quantum mechanical calculations for $N_{\rm QM}(\varepsilon)$ and its
semiclassical counterpart $N_{BS}(\varepsilon)$, which are
indicated by the shaded areas in Fig.~\ref{fig2}(a) and (b). For the two
systems considered in Fig.~\ref{fig2} we note that the mismatch
between the quantum and the semiclassical results is located at rather
well-defined values of the excitation energy $\varepsilon$. 
Note that the mismatch tends to occur at
values of $\varepsilon$ just {\it above} a cusp.

Since pathlength and energy at fixed quantum number $n$ are inversely
proportional to each other [Eq.~(\ref{epsqn})] the energies {\it above} the cusp
which is associated to quantum number $n$ [e.g.~the one marked $I$ 
in Fig.~\ref{fig2}(b)] correspond to the longest orbits associated
with quantum number $n+1$. The deficiencies of the BS approximation are
evidently caused by contributions from long orbits. This observation is in
line with the well-known failure of the BS approximation to yield the energy
gap for chaotic $N$ billiards,\cite{Vavilov,Altland,Melsen} also 
caused by long orbits. We also see a discrepancy
above the cusp marked $II$ in Fig.~\ref{fig2}(b) which turns out to be
due to diffractive
scattering at the corner. We will discuss diffractive effects in 
more detail below when we explore different mechanisms 
for the failure of the BS approximation by changing the geometry.

\subsection{Stretched separable billiard}

\begin{figure}[!t]
\epsfig{file = 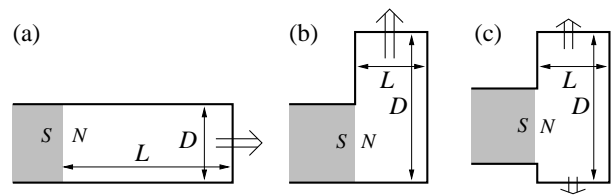, width = 8cm}
\caption{Geometries with tunable boundaries: 
  (a) stretched N-cavities with $L\gg D$ but with $D$ 
equal to the width of the SN-interface $W=D$. 
(b) stretched N-cavities with $D \gg L$ but constant length $L$ 
(lower boundary of S and N aligned). (c) as (b) but SN-interface 
at arbitrary position on the entrance side.}
\label{fig2geom}
\end{figure}

\begin{figure}[h]
\epsfig{file=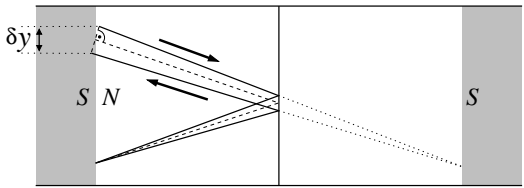,width=6.9cm}
\caption{Imperfect Andreev reflection in the fundamental and 
extended zone scheme. The returning hole hits 
the SN-interface a distance $\dely$ apart from the starting point of 
the particle.}
\label{figdely}
\end{figure}

We consider now the separable case $W=D$ but stretch the billiard to $D/L\ll
1$ [see Fig.~\ref{fig2geom}(a)] without, however, 
changing the Fermi energy $E_F$ 
of the system. The distance $l$ travelled between two encounters with the
SN-interface is, for the shortest orbits, at least $2L$. The effect
of imperfect retracing, originating 
from the difference in angles between $\theta_e$ and
$\theta_h$, is a divergence between electron
and hole trajectories ($\propto l$),
which is amplified by large L. We introduce as a measure for the
imperfect retracing the lateral displacement $\delta y$ on the interface
between the hole and the particle trajectory after one loop (see Fig.~\ref{figdely}),

\begin{eqnarray}
  \dely &=& l \left|\sin(\theta_e) - \sin(\theta_h)\right| \nonumber \\
         &=& l \frac{\varepsilon\sin\theta}{\EF} + 
O(\varepsilon^2)\,, \label{dely}
\end{eqnarray} 
where $\theta =\theta_{e,h}|_{\varepsilon=0}$. For
the particular geometry considered [Fig.~\ref{fig2geom}(a)], $\dely$
increases linearly with both the trajectory length 
$l$ and the excitation energy $\varepsilon$. Note that the ratio 
$\varepsilon/\EF$ is, in general, much smaller than one, 
i.e.~$\varepsilon\le\Delta \ll \EF$.

This classical scale for the mismatch between the particle and 
hole orbits should
be related to the quantum scale, i.e.~the linear dimension of the wave packet
estimated by the de Broglie wavelength $\lambda_F$. We thus introduce
$r=|\delta y|/\lambda_F$ as order parameter for the error of retracing. For
$r\ll 1$ the wave packet cannot resolve the mismatch and the BS approximation
should work well. Conversely, as $r$ reaches the order of 
unity, quantization based on the
existence of a continuum of periodic orbits should fail.

\begin{figure}[!b]
\epsfig{file=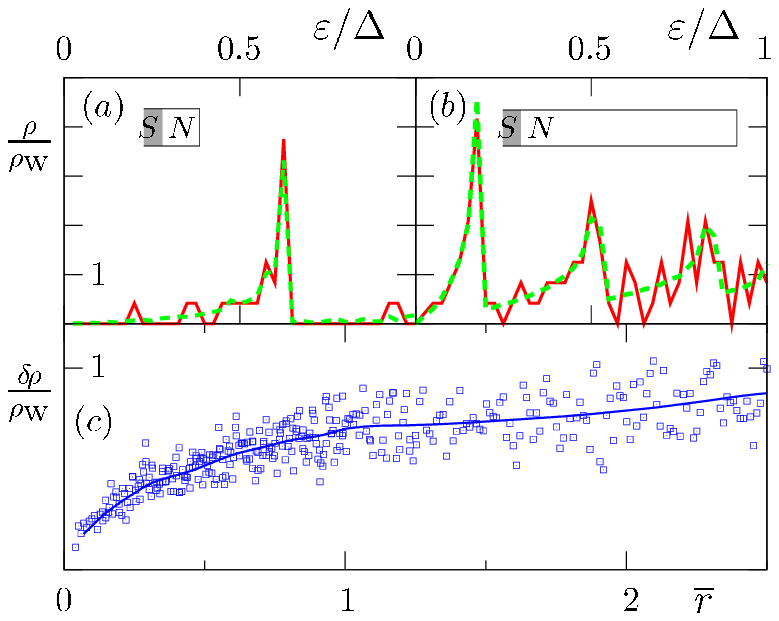,width=8.5cm}
\caption{(Color online) The DOS calculated quantum mechanically 
(red solid curve) and by the BS approximation
(green dashed curve) for two different geometries with ratios 
of (a) $L/D = 1$  and (b) $L/D=6$ (see insets) in units of the 
Weyl approximation of the DOS
per unit area, $\rho_W = m_{\rm eff}/(\hbar^2\pi)$. 
(c) error of the semiclassical prediction 
$\delN$ [Eq.~(\ref{small:delN})] as a function of an energy-averaged 
$\overline r$
[Eq.~(\ref{delyAv})] for 300 different ratios of $L/D \in [1,20]$\,.  
The solid blue line shows a smoothed average (30 adjacent points)
of the recorded data.
}
\label{figWdY}
\end{figure}

To probe the breakdown of the retracing approximation quantitatively, we
compare the semiclassical with the quantum density of states (DOS), which are
obtained from the state counting function $N(\varepsilon)$ as 
\begin{equation}\label{DOS}
\rho_{\rm{BS,QM}}(\varepsilon) = \frac{\partial N_{\rm
    BS,QM}(\varepsilon)}{\partial\varepsilon}\,,
\end{equation}
since the DOS is more sensitive to errors than the (smoothed) spectral staircase.
Results for two cavities with different values of $L$ are shown in
Fig.~\ref{figWdY}(a) and (b). This example illustrates that the degree of
agreement between
$\rho_{\rm{QM}}$ and $\rho_{\rm{BS}}$ is indeed controlled by $r$. 
The following trends can be observed: 
(i) Due to the comparatively shorter length $l$ of
orbits in (a), the overall agreement there is better in (a) than in (b). 
(ii) The agreement in (b) deteriorates for larger values of
$\varepsilon/\Delta$,
since the mismatch in retracing increases with $\varepsilon$.

As a measure for the average mismatch between $\rho_{\rm QM}$ and $\rho_{\rm BS}$ 
in an Andreev billiard of given
geometry we use the quantity $\delyAv$, which is $\dely$ averaged 
over all $\varepsilon$ and all angles,
\begin{equation}
   \delyAv = \frac{1}{\pi\Delta}\int_{-\pi/2}^{\pi/2} 
d\theta\,\cos\theta  \int_0^\Delta d\varepsilon \,\dely 
= \frac{2L\Delta}{\pi\EF}\,,\label{delyAv}
\end{equation}
since for this particular geometry $l=2L/\cos\theta\,.$ The error in the DOS is
quantified in terms of the root mean
square (RMS) deviation $\delN$ 
between $\rho_{\rm{BS}}$ and $\rho_{\rm{QM}}$, 
\begin{equation}
  \delN  = \sqrt{\frac{1}{\Delta}\int_0^\Delta {\rm{d}} \varepsilon \;\left|
  \rho_{\rm QM}-\rho_{\rm BS}
\right|^2}\,.\label{small:delN}
\end{equation}
As expected, as long as $\overline r = \delyAv/\lamF \ll 1$, 
the retracing approximation is sufficiently
accurate to reproduce the DOS on the scale of the mean level spacing 
[see Fig.~\ref{figWdY}(c)]. Note
that resolution of individual levels is beyond the scope of the BS
approximation. As discussed below, an EBK quantization is needed for an
accurate quantization of individual levels. As $\overline r$ 
approaches unity the
oscillations in the quantum DOS cannot
be resolved any longer. The error $\delN(\delyAv)$ appears to saturate, in
agreement with the observation that in this regime the main peaks in 
Fig.~\ref{figWdY}(b) [corresponding to the main cusps in $N(\varepsilon)$]
remain well described by the retracing approximation.
Identifying the latter with the shortest Andreev-reflected orbits of length
$\sim 2L$ allows a simple explanation as to why even a drastic elongation of
the cavity leaves these cusps well described by the retracing approximation:
orbits of length $\sim 2L$ correspond to strictly horizontally 
injected trajectories
(with $\theta=0$). They do not accumulate any lateral displacement 
$\dely$ irrespective of the length of the orbit and $\varepsilon$. While the
average value $\delyAv$ is increasing with growing cavity elongation
(due to orbits with $\theta\neq 0$), the horizontal orbits limit the increase
$\delN(\delyAv)$, leading to the observed saturation in Fig.~\ref{figWdY}(c)
for $\delyAv > \lamF$.

\begin{figure}[!t]
\epsfig{file=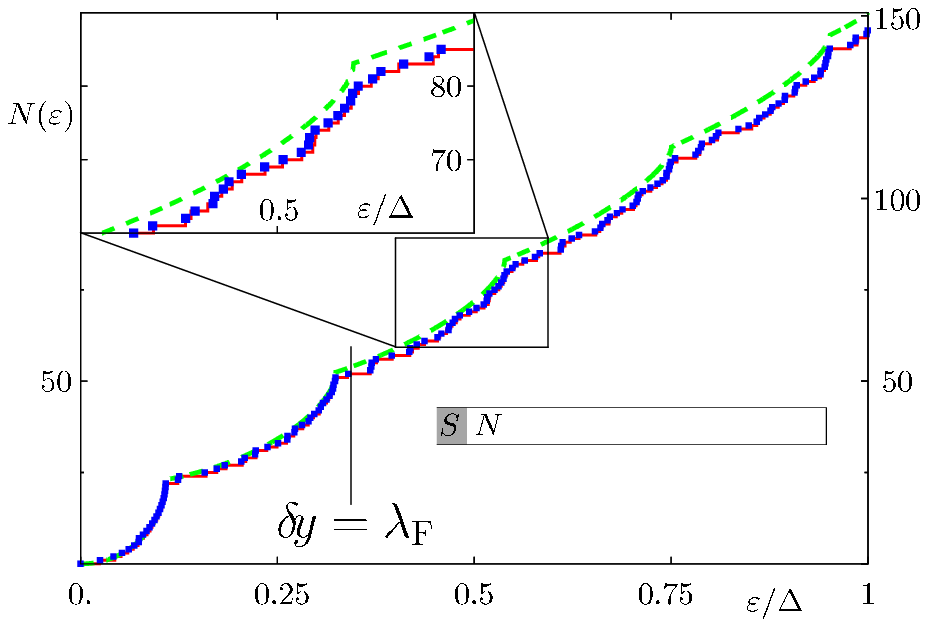,width=8.5cm}
\caption{(Color online) Comparison of the exact quantum mechanical 
state counting function
  $N_{\rm QM}$ (red staircase) with 
the semiclassical BS (green dashed line) and EBK (blue squares)
approximations for a
highly elongated cavity ($L/D = 10$), as shown in the lower inset. Quantum
mechanical and EBK solutions are nearly indistinguishable. The
upper inset shows a magnification. 
The energy for which $\dely=\lamF$ is marked.}
\label{figEBK}
\end{figure}

Additional evidence that the discrepancy between the quantum DOS and the BS
spectrum is due to the retracing assumption (A1) and not due to the failure of
semiclassics itself (A3) can be gained from an EBK quantization. Note that the
Andreev system [Fig.~\ref{fig2geom}(a)] 
is separable. Moreover, regions in phase space that
will not make contact with the SN-interface are in this geometry 
of measure zero,
i.e.~assumption A2 is valid. In sharp contrast to the discrepancy between
$N_{\rm QM}$  and $N_{\rm BS}$, we observe that $N_{\rm QM}$  and $N_{\rm EBK}$
agree almost perfectly in the whole energy interval $0\le\varepsilon\le\Delta$ 
(see Fig.~\ref{figEBK})
even for very elongated cavities where most orbits are non-retracing ($\delta
y>\lambda_F$). The criterion $\delta y\approx\lambda_F$ is found
to be a good estimate where the BS quantization deviates from both the EBK
quantization and the exact quantum spectrum. The case of the elongated
$N$-billiard clearly demonstrates that the failure of the BS quantization 
in this case is
due to the retracing assumption (A1), but 
not due to the failure of semiclassics (A3) itself. 

\subsection{Stretched non-separable billiard}

\begin{figure}[htb]
\epsfig{file=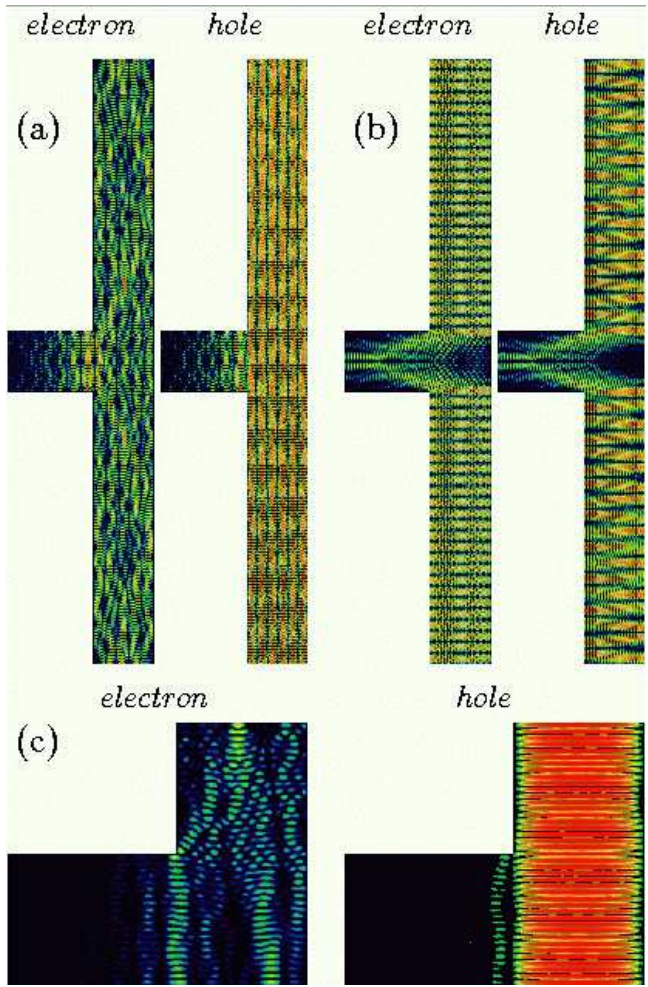,width=8.5cm}
\caption{(Color online) Selected Andreev eigenstates of non-separable Andreev
billiards stretched parallel to the SN-interface displaying a large 
discrepancy between electron and hole wavefunction patterns.
(a) $L/D=10$, $\varepsilon=0.760\Delta$; 
(b) $L/D=10$, $\varepsilon=0.766\Delta$, $\kf=21.5\pi/W$;
(c) $L/D=2$, $\varepsilon=0.040\Delta$, $\kf=20.5\pi/W$;
}
\label{figLong}
\end{figure}

\begin{figure}[htb]
\epsfig{file=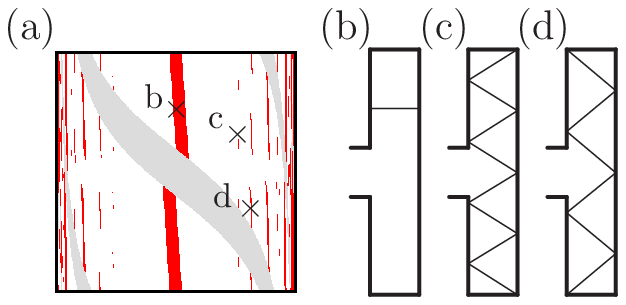,width=8.5cm}
\caption{(Color online) (a) Poincar\'e 
surface of section of a stretched non-separable Andreev 
billiard, taken at the right vertical side, opposing the SN-interface:
$W/D=5$, $W=L$.
The light gray area marks regions of classical phase space coupled directly
to the entrance lead, with one single reflection at the right wall. 
Dark red areas show regions of classical phase space decoupled from the
superconducting lead. Classical trajectories corresponding
to selected areas of phase space are shown in (b)-(d).}
\label{figPhase}
\end{figure}

We focus now on billiards stretched in the direction of the SN-interface with
$D/W\gg 1$ and $D/L\gg 1$ [Fig.~\ref{fig2geom}(b,c)]. 
For $D/W > 1$ the Andreev billiard
becomes non-separable and an EBK quantization is no longer rigorously
justified. However, as we will show below, the predictions by the BS
quantization can still be compared with an approximate EBK quantization. As
the wave functions (Fig.~\ref{figLong}) 
clearly indicate, the retracing approximation
breaks down with large discrepancies between the particle ($|u|^2$) and hole
($|v|^2$) densities occurring. 
For this system, one obvious culprit is assumption
A2. An extended region of classical phase space (Fig.~\ref{figPhase}) 
remains decoupled from the SN-interface. 
Consequently, the quantum DOS is associated, in part, with decoupled
regions which are not represented at all by the BS approximation. 
It has previously been pointed out for other geometries (whispering gallery
trajectories in circular billiards) 
that Andreev billiards indeed feature quantum states that are not
associated with periodic orbits.\cite{JozBoxDisk}

A more detailed inspection reveals that a certain class of wave functions 
has the following features: (1) 
The probability density at the SN-interface is typically low --
corresponding to decoupling from the superconductor. This suggests that the
SN-interface effectively acts like a hard-wall boundary. (2)
Integrating the probability density in the electron- and hole-part of the
N-region typically shows a strong asymmetry. For the case
depicted in Fig.~\ref{figLong}(c) we calculate a probability of $98.3\%$ 
for the hole component and only $1.3\%$ for the particle component. 
The coupling with the superconductor is very weak, resulting
in a probability of $0.4\%$ of finding the particle in the 
superconducting lead. These
observations suggest that the particles in such non-retracing states are
quasibound in either the electron- or the hole-space with only infrequent
and short excursions to their respective mirror-space. As a consequence we may
employ again an EBK quantization scheme to estimate the corresponding energy
level of the Andreev state, however now in the space of the quasi-bound
particle or hole. Exploiting the assumption of weak coupling to the
SN-interface, we replace the SN-interface by a hard-wall (box) boundary. 
Applying this approach e.g.~to the state depicted in 
Fig.~\ref{figLong}(a), the EBK
quantization  
\begin{equation}
E_{n,m} = \frac{1}{2}\left(\frac{n\pi}{L}\right)^2 +
\frac{1}{2}\left(\frac{m\pi}{D}\right)^2\,
\end{equation}
implies for the density of the ``isolated'' electron state
$n=6$ maxima in $x$-direction and $m=205$ maxima in $y$-direction.
Inserting these quantum numbers yields an excitation energy of
$\varepsilon_{\rm box} = 0.753\Delta$, which
is very close to the exact eigenenergy of the Andreev eigenstate of
$\varepsilon_{\rm QM} = 0.760\Delta$. Analogously, the state depicted in
Fig.~\ref{figLong}(c) has quantum numbers $n=1$, $m = 41$, corresponding
to a hole excitation of $\varepsilon_{\rm box} = 0.038\Delta$, which compares
well to the eigenenergy of the Andreev eigenstate of 
$\varepsilon_{\rm QM} = 0.040\Delta$.
While the position of these Andreev states can thus be explained fairly
accurately by a box (or EBK) quantization, they are obviously not included in a 
standard BS approximation which considers only retracing electron-hole
orbits. This illustrates the failure of assumption A2, while 
semiclassics still remains applicable. 

\subsection{Diffractive effects}
\label{secDiff}
\begin{figure}
\epsfig{file=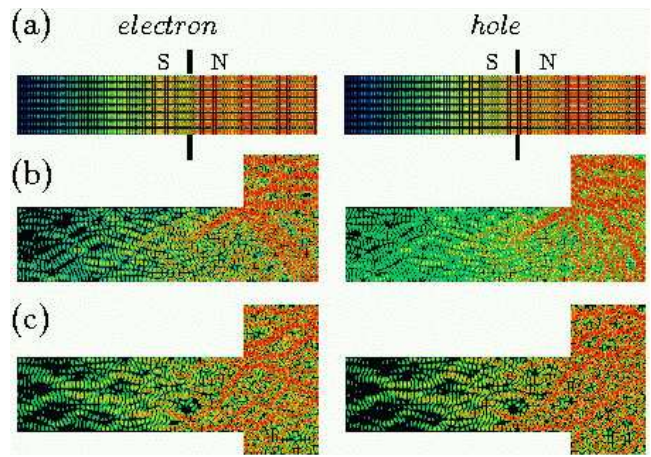,width=8.5cm,clip=true}
\caption{(Color online) Selected Andreev eigenstates of systems with 
(a) no diffractive corners 
($L/D=1.6$, $W=D$, $\kf = 20.5\pi/W$, $\varepsilon/\Delta=0.48$);
(b) one diffractive corner 
($D/L=1.7$, $W=L$, $\kf = 21.5\pi/W$, $\varepsilon/\Delta=0.72$); 
(c) two diffractive corners
($D/L=1.7$, $W=L$, $\kf = 20.5\pi/W$, $\varepsilon/\Delta=0.82$).} 
\label{figDiff}
\end{figure}

For the present hard-wall $N$-cavities, introducing non-separability implies
simultaneously the introduction of diffractive edges and corners
 (see Fig.~\ref{fig2geom}).\cite{Wiersig2}
In the preceeding example, diffractive effects were present but for
the special group of states considered of minor importance. The latter was
confirmed not only by the density distributions in the Andreev states
(Fig.~\ref{figLong}) but also by the accuracy 
of the EBK prediction. We can now enhance
diffractive effects by considering only moderately stretched cavities $D/W>1$
with arbitrary position of the SN-interface. As shown in Fig.~\ref{figDiff}, 
we can thus realize Andreev
billiards with zero corners ($D/W=1$), one corner 
($D/W>1$, lower boundary of S and N aligned), 
and two corners 
($D/W>1$, arbitrary position of the SN-interface).
Diffractive effects originate from potential variations on a length scale
smaller than the de Broglie wavelength. In a semiclassical picture,
diffractive scattering gives rise to a non-deterministic, stochastic angular
distribution. It certainly prevents the deterministic perfect 
retracing of orbits.

Diffractive scattering can be directly visualized in the wave function
[Fig.~\ref{figDiff}(b,c)]. 
The density enhancement representing a scar of a classical
trajectory hits the corner of the SN-interface. As a result the trajectory
``splits'' into a bouncing-ball orbit and a broad angular
distribution. Significant differences between the particle ($|u|^2$) and 
hole ($|v|^2$) density in the retracing orbit are clearly visible.

\begin{figure}[htb]
\epsfig{file=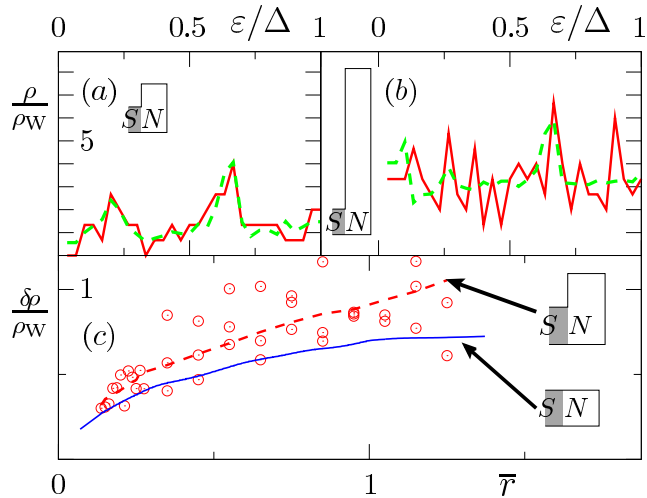,width=8.5cm}
\caption{(Color online) The DOS calculated quantum mechanically 
(red solid curve) and by the BS approximation
(green dashed curve) for two different geometries with ratios 
of (a) $L/D = 1$ and (b) $1/8$ (see insets), 
in units of the Weyl approximation of the DOS
per unit area, $\rho_W = m_{\rm eff}/(\hbar^2\pi)$.
(c) Error of the retracing approximation $\delN$ [Eq.~(\ref{small:delN})] for
transversely elongated cavities (red circles)
as a function of $\overline r$.
The solid blue (dashed red) line shows the average 
error for longitudinally (transversely) elongated cavities 
(see also Fig.~\ref{figWdY}).}
\label{figWdWy}
\end{figure}

For a more quantitative assessment we compare again the
quantum and the BS-results for the DOS [see Fig.~\ref{figWdWy}(a,b)] 
and find, similarly to the case of the separable geometry,
that for larger $D$ the mismatch $\delta\rho$ in the DOS increases. If we 
compare, however, how the error $\delta\rho$ scales with the geometric
retracing mismatch $\delta y$ [Fig.~\ref{figWdWy}(c)], 
we find that in the Andreev billiard with a
diffractive corner the average mismatch 
is considerably larger than for the geometry without a sharp corner. 
The difference in $\delta\rho$ at fixed $\delyAv$ can be thus attributed to
diffraction. Note that diffractive scattering is an additional mechanism for
the failure of the retracing approximation, which is independent
of the failure induced by long electron-hole orbits. 

\begin{figure}[!b]
\epsfig{file=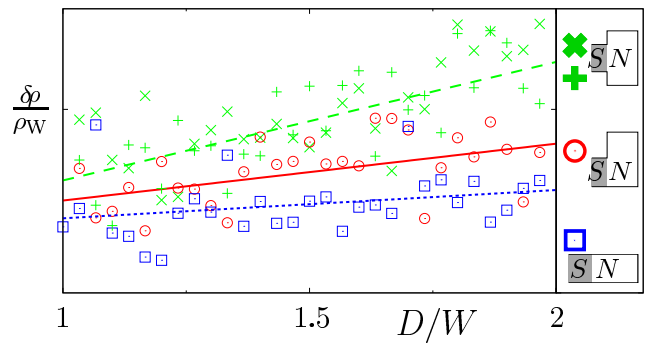,width=8.5cm}
\caption{(Color online) Error of the retracing approximation $\delta\rho$ as a
  function of $D/W$ for geometries with no (blue square, dashed line), one 
(red circle, solid line)
  and two (green cross, dashed line) diffractive corners. Different orientations
  of the cross refer to different positions of the superconducting lead.
  The lines show a linear root mean square fit of the data. 
}
\label{figDosCorners}
\end{figure}

Additional evidence along these lines is presented in 
Fig.~\ref{figDosCorners} where the mismatch $\delta\rho$
between the BS approximation and the quantum results as a function of the
geometrical aspect ratio of the normal-conducting region is compared for
three geometries with either (a) zero, (b) one, or (c) two
diffractive corners at the SN-interface. As expected, the 
mismatch increases (1) for larger aspect ratios and (2) at fixed aspect ratio
for an increasing number of diffractive corners. Further support 
comes from the observation that the
mismatch $\delta \rho$ for the cavity with two diffractive corners is, to a
large degree, independent of the position of the lead, as shown in
Fig.~\ref{figDosCorners}.

Unlike other sources for the failure of the retracing approximation discussed
above, diffraction limits the validity of standard semiclassical
approximations itself, i.e.~of assumption A3. Methods for describing 
diffractive effects semiclassically by including 
pseudopaths\cite{Stampfer,Stampfer2}
and non-deterministic scattering\cite{Aigner} have been explored for
scattering at open $N$ cavities. They play a crucial role in determining the
$S$-matrix for such devices. The present results demonstrate that 
diffractive corrections come into play also for Andreev systems.

\section{conclusion}\label{SecSum}
We study Andreev billiards of rectangular shape and compare 
the quantum solutions of the Bogoliubov-de Gennes equation with a
semiclassical Bohr-Sommerfeld approximation that assumes 
retracing electron-hole orbits as a result of perfect Andreev reflection. 
Investigating 
the validity of this widely used semiclassical approach,
we identify three independent mechanisms which influence the density of bound
Andreev states beyond the simple BS quantization: (i) long trajectories
in the billiard typically magnify deviations between the electron and the
hole part of orbits and cause a failure of the retracing approximation.
This is of importance for the understanding of the failure of the BS 
approximation to reproduce the hard gap for chaotic N cavities. The latter is controlled
by the behavior of long paths for which, as we show, the retracing approximation breaks down.
(ii) BS quantization fails when a subspace of the phase space is decoupled
from the SN-interface and thus inaccessible to retracing orbits. (iii)
diffractive effects limit the validity of the semiclassical quantization. In
the latter case the failure is not just due to the assumption of retracing,
but due to the failure of standard semiclassics itself. In this case, 
also alternative
methods such as EBK quantization of separable or 
adiabatic quantization
\cite{Silvestrov,Goorden}
 of (smooth) chaotic systems will fail. Corrections beyond
standard semiclassics, e.g.~inclusion of pseudo-paths\cite{Stampfer,Stampfer2} of 
diffracted orbits will be needed. 
\begin{acknowledgments}
We thank F.~Aigner, J.~Cserti, B.~L.~Gy\"orffy, A.~Korm\'anyos, and C.~J.~Lambert 
for valuable discussions.
Support by the Austrian Science Foundation (Grant No.~FWF-P17359), the 
Hungarian-Austrian Intergovermental S\&T cooperation program (Project
No. 2/2003) and the British Council Vienna
is gratefully acknowledged. 
\end{acknowledgments}

\end{document}